\documentstyle[12pt,tighten,eqsecnum,epsf,aps]{revtex}

\begin{document}
\draft

\title{$g_{\omega\rho\pi}$ reexamined}
\author{M. Lublinsky\thanks{E-mail: mal@techunix.technion.ac.il}}
\address{Department of Physics, Technion -- Israel Institute of 
Technology, Haifa 32000, Israel}
 \maketitle

\begin{abstract}
The transition constant for the hadronic decay $\omega \rightarrow \rho\pi$
is investigated by means of QCD sum rules in external axial field. 
The obtained value for $g_{\omega\rho\pi}$ is about 16 GeV${}^{-1}$ that
is in a good agreement with experimental data.
\end{abstract}


\section{Introduction}
\label{sec:intr}

During  last several years, there was a  significant increase in the 
interest  in the low energy QCD and in understanding the low energy hadron 
dynamics. This happened both due to the theoretical progress and to the
availability of  new experimental data. However, the problem of 
understanding the low energy properties of hadrons is still far from being
solved, due to the need to understand the nonperturbative QCD effects that
govern the hadron dynamics.

The QCD sum rules, first suggested by  Shifman, Vainshtein and  Zakharov 
\cite{SVZ79} (see also Ref. \cite{RRY85} for an extensive review), 
historically are one of the most effective tools used to study the low energy 
properties of hadrons. The method was successfully applied to numerous low 
energy problems, such as the determination of the mass spectrum of the 
low-lying resonances, of the form-factors etc.  

The goal of the present note is to use QCD sum rules in order to calculate
the $\omega \rightarrow \rho\pi$ transition constant. Let us recall  that
this constant is defined  \cite{EK83} by the equation:
\begin{equation}
M_{\omega \rightarrow \rho\pi}=g_{\omega\rho\pi} {\epsilon}_{ij\alpha\beta}
p^{i} p \prime ^{j} e_{\omega}^{\alpha} e_{\rho}^{\beta}
 \label{h1}
\end{equation} 
Here $M$ is the invariant amplitude for the $\omega \rightarrow \rho\pi$ 
transition,
$p, e_{\omega}$ and $p \prime ,e_{\rho}$ denote  momenta and 
polarisations of \  $\omega$ \ and \ $\rho$. Experimental \ value for \ 
$g_{\omega\rho\pi}$ \  is \ about \ 14 $\pm$ 2 Gev ${}^{-1}$. 
Within the framework of QCD sum rules, $g_{\omega\rho\pi}$ was already
considered several times \cite {EK83,RRY83,NP84,Kh85,MS88,BF89}. The
results obtained in Ref. \cite{EK83,RRY83,NP84} varied from 17 Gev${}^{-1}$
to 15 Gev ${}^{-1}$, that were in an agreement with the experimental data
up to 20\% - 30\% - the usual error of the method. Ref \cite{Kh85} gives for
this constant 9 Gev ${}^{-1}$ that contradicts both the previous calculations
and the experimental data. The authors of Ref. \cite{MS88} present two 
results. The first one supports calculations of Ref. \cite{Kh85}. In order to
obtain the second result they modified sum rules incorporating the ABJ axial
anomaly and finally $g_{\omega\rho\pi}$ was estimated as 16 Gev ${}^{-1}$.
In Ref. \cite{BF89}, the light cone sum rule method was used to determine
a wave-function of $\pi$ meson. The results gave 
15 Gev ${}^{-1}$  for $g_{\omega\rho\pi}$. Some 
uncertainty in predictions and a significant mismatch between results of 
Ref. \cite{Kh85} and the other obtained results have drawn us to reconsider 
$g_{\omega\rho\pi}$.  

In this note we calculate $g_{\omega\rho\pi}$ using the QCD sum rules for 
the polarisation operator in  external axial field.  This method was 
first discovered in Ref \cite{IS83} and since was used to determine the
magnetic moments of baryons \cite{IS83}, axial coupling constants and
electro-magnetic form-factors of nucleons \cite{BK83,NR84,BK93} at low 
momenta transfer, to study $D^0 \rightarrow D\gamma,Dn$ transitions
\cite{EK85}, etc.

The main advantage of this method is that it permits one to work directly
in the  region of small momenta transfer (by paying the price of  
introducing new condensates, see below). Note that this is exactly the case
at hand: the masses of  $\omega$ and $\rho$ are almost equal, hence the
momentum of virtual $\pi$ meson is  expected to be small.

The structure of the paper is as follows. In Sec. 2 we shortly remind the 
 reader the basics of QCD sum rules method in 
external field. Sec. 3 is devoted to the main calculations 
and  in the  Sec. 4 we give the conclusions.  

\section{external field method}
\label{sec:method}

Let us review in short the QCD sum rules for polarisation operator in 
external  field.

We start with the polarisation operator:
\begin{equation}
\Pi ^{\mu \nu}_A(p,q)= - \int d^4 x e^{-ipx}\langle j_{\omega}^{\mu}(x)
j_{\rho}^{\nu}(0)\rangle _A
\label{h3}
\end{equation}
Here $j_{\omega}^{\mu},j_{\rho}^{\nu}$ are the quark currents that create 
$\omega ^0$ and $\rho ^0$ mesons: 
\begin{eqnarray}
& & j_{\omega}^{\mu}= \frac{1}{6}(\bar{u}{\gamma}^{\mu}u+
	\bar{d}{\gamma}^{\mu}d) \nonumber \\ \label{h2} \\
& & j_{\rho}^{\nu}= \frac{1}{2}(\bar{u}{\gamma}^{\nu}u-
\bar{d}{\gamma}^{\nu}d)  \nonumber
\end{eqnarray}

The index $A$ means that we consider this correlator in the external 
axial field
$A(x)=Ae^{-iqx}$ that corresponds to the following term in the Lagrangian:

\begin{equation}
L_A=-\frac{1}{2} \int d^4 y e^{-iqy}j_{5}^{\lambda}A_{\lambda}
\label{h4}
\end{equation}
Here $q$ is the external field momentum and $j_{5}^{\lambda}$ is the
quark current that creates pion:
\begin{equation}
j_{5}^{\lambda}= \frac{1}{2}(\bar{u}{\gamma}^{\lambda}{\gamma}_{5}u-
	\bar{d}{\gamma}^{\lambda}{\gamma}_{5}d)
\label{hh2}
\end{equation}
The expression (\ref{h3}) contains, generally speaking, terms of different
powers in $A_\alpha$. We shall be interested here in the terms linear in $A$
only. They appear both  when considering the propagator of the quarks in the
external field  and when taking into account the interaction of the external 
field  with the quark condensate.  Moreover, as it is explained below, 
it is enough to carry the actual calculations for the case $q_\alpha=0$, i.e.
for the case of constant external field.  
In the latter case, in the approximation linear in $A_\alpha$, the quark 
propagator in the $x$ representation is the following:
\begin{equation}
S_A(x)=-\frac{\hat{x}}{2\pi ^2 x^4}-i\frac{(Ax)\hat{x}\gamma _5}{2\pi ^2 x^4}
\label{h7}
\end{equation}
The terms in the operator expansion which correspond to the nonperturbative
interaction of the quarks with the axial field have been obtained in Ref.
\cite{BK83} and have the following form:
\begin{eqnarray}
\langle :\bar{\psi}\psi :\rangle _A & = & 
\delta ^{ab}\frac{f_{\pi}^2}{12}\hat{A}\gamma_5 +
i\delta ^{ab}\frac{\langle 
:\bar{\psi}\psi :\rangle }{3^2 2^3}(\hat{x}\hat{A}-
\hat{A}\hat{x})\gamma_5 \nonumber \\
& - & \delta ^{ab}m_1^2\frac{f_{\pi}^2}{6^3}(\frac{5}{2}
x^3(A\gamma_5)-(Ax)\hat{x}\gamma_5)
\label{h8}
\end{eqnarray}
Here, \ $f_{\pi}=135$ Mev,
\ $\langle :\bar{\psi}\psi :\rangle =-(240$ Mev)$^3$, \ 
$m_1^2=0.2$~Gev${}^2$ is the parameter
of the matrix element:
\begin{eqnarray}
\langle 0|g_s\bar{u}\tilde{G}_{\mu\nu}^i(\lambda^i/2)
\gamma_{\nu}d|\pi\rangle =im_1^2f_{\pi}
q_\mu \nonumber
\end{eqnarray}
which is defined in Ref. \cite{NSVZ84}, and $\tilde{G}_{\mu\nu}^i$ is the
dual tensor of the gluon field strength.
By means of expressions (\ref{h7}), (\ref{h8}), and using the technique
developed in Ref. \cite{IS83,BK83}, we are able to determine the
 polarisation operator
$ \Pi ^{\mu \nu}_A(p,q)$ in the region  $p\prime^2, p^2\sim -1$ Gev$^2$.
The actual calculation is carried through using the Wilson operator product
expansion (OPE) and taking into account several first low dimensional
operators:
\begin{eqnarray}
\gamma_\mu(x)\gamma_\nu(0)=\sum C_n(x^2)O_n(0), \nonumber
\end{eqnarray}
where $C_n(x^2)$ are coefficient functions.

In order 
to extract pion pole from the amplitude (\ref{h1}) we have to manage  the
case of small, but nonzero $q$. It was suggested in Ref. \cite{EK85} that in
 the case of small but nonzero $q$, one may perform the 
calculation of the polarisation operator for $q=0$ and then, in the final
expression, make the substitution:
\begin{equation}
A_\alpha \rightarrow ({g_{\alpha}}^{\lambda}-\frac{q_{\alpha}q^{\lambda}}
{q^2})A_{\lambda}
\label{h9}
\end{equation}
The calculation of the polarisation operator (\ref{h3}) by means of OPE
gives the right-hand, theoretical part of the sum rule. On the other side,
the polarisation operator satisfies the dispersion relations, which we shall
saturate by the contribution of the intermediate states. The dispersion 
relations are:  
\begin{eqnarray}
f(Q^2)&=&\int_0^\infty {\frac{\rho(s)ds}{(s+Q^2)^2}}+
\int_0^\infty {\frac{\rho_1(s)ds}{s+Q^2}} \nonumber\\ 
& &Q^2=-p^2  
\label{hh3}
\end{eqnarray}
where $\rho(s)$ is the contribution in (\ref{h3}) of the diagonal transitions
corresponding to the diagram (Fig.\ \ref{fig:gr1}), $\rho_1(s)$ is the 
contribution
of the non-diagonal transitions (Fig.\ \ref{fig:gr2}). $f(Q^2)$ 
denotes the structure function of the polarisation operator 
$\Pi^{\mu\nu}_A$
that we use  for the calculation of $g_{\omega\rho\pi}$.

After saturating (\ref{hh3}) by resonances, taking into account $\omega,\rho$
resonance contributions explicitly, and taking care of higher exited states
using continuum model \cite{I81}, we obtain the phenomenological, left-hand
part of the sum rule.
\section{sum rule}
\label{sec:sumr}
Apply now the ideas discussed above to the 
$\omega\rightarrow\rho\pi$ transition. 

To begin with, consider the phenomenological part. We saturate the
currents (\ref{h2}), (\ref{hh2}) by the physical resonances and define 
the residues:
\begin{eqnarray}
& & \langle 0|j_{\omega}^{\mu}|\omega\rangle 
=\frac{m_{\omega}^2}{g_{\omega}}
       e_{\omega}^{\mu} \nonumber \\
& & \langle 0|j_{\rho}^{\nu}|\rho\rangle =
\frac{m_{\rho}^2}{g_{\rho}}e_{\rho}^{\nu}
       \label{h11} \\
& & \langle 0|j_{5}^{\lambda}|\pi\rangle 
=-i\frac{f_{\pi}}{\sqrt 2}q^{\lambda} \nonumber
\end{eqnarray}
Here, $m_{\omega}, g_{\omega}$ and  $m_{\rho}, g_{\rho}$ denote masses and
form-factors of $\omega$ and $\rho$ mesons. $f_{\pi}$ and $q$ are the 
 form-factor of $\pi$ meson in chiral limit  and its momentum. 

The linear (in $A$) part  of the polarisation operator 
$\Pi ^{\mu \nu}_A(p,q)$ is
 proportional to the 
invariant amplitude (\ref{h1}) and is given by:
\begin{equation}
\Pi ^{\mu \nu}_A(p,q)=M\frac{\langle 0|j_{5}^{\lambda}|\pi\rangle 
A_{\lambda}}{q^2}
	\frac{\langle 0|j_{\omega}^{\mu}|\omega\rangle }{(p^2-m_{\omega}^2)}
	\frac{\langle 0|j_{\rho}^{\nu}|\rho\rangle }{({p}\prime ^2
-m_{\rho}^2)}+...
\label{h12}
\end{equation}
\begin{eqnarray}
p\prime =p+q \nonumber
\end{eqnarray}
The ... denote contributions of other exited states and ``parasitic'' 
non-diagonal pieces.
Taking limit $q \rightarrow 0$ and substituting (\ref{h1}) and (\ref{h11}),
into (\ref{h12}) we arrive to the expression for $\Pi ^{\mu \nu}(p,q)$:
\begin{equation}
\Pi ^{\mu \nu}_A(p,q)=-ig_{\omega\rho\pi}\frac{m_{\rho}^2 m_{\omega}^2}
{\sqrt 2g_{\rho}
g_{\omega}} \frac{f_{\pi}}{(p^2-m_{\rho}^2)^2}\frac{q^{\lambda}A_{\lambda}}
{q^2}\epsilon_{\mu\nu\alpha\beta}q^{\alpha}p^{\beta}
\label{h13}
\end{equation}
In order to obtain (\ref{h13}) we used the well-known expression 
\begin{eqnarray}
\langle e^\alpha e^\beta\rangle =g^{\alpha\beta}-\frac{p^\alpha 
p^\beta}{p^2} \nonumber
\end{eqnarray}
for the density matrix, we took into account $m_{\omega}\approx m_{\rho}$, 
and we have written explicitly the contribution of the
$\omega\rightarrow\rho$ transition only.
We need to consider the sum rule for the following structure:
\begin{equation}
\frac{q^{\lambda}A_{\lambda}}
{q^2}\epsilon_{\mu\nu\alpha\beta}q^{\alpha}p^{\beta} 
\label{h14}
\end{equation}
In the subsequent calculations of OPE we keep only the terms that  contain
the structure (\ref{h14}).
\vskip10pt

Let us now turn to the left, theoretical part. We restrict ourselves 
to the  operators of dimensions $d\le 5$ only. \\
Consider the unit operator, first. 
Its contribution is given by one diagram of  zero order in 
$\alpha_s$ (Fig.\ \ref{fig:gr3}) and six diagrams of the first order in 
$\alpha_s$ (Fig.\ \ref{fig:gr4}). 

The explicit calculation of the expression corresponding to the 
(Fig.\ \ref{fig:gr3}) gives zero. 

Since the contribution of zero order to the unit operator $I$ vanishes,
 the $\alpha_s$ 
corrections  to this operator are taken into account. We shall do it because
the contribution of this operator is potentially the leading one and even
 $\alpha_s$ corrections due to the unit operator may give large 
contribution to the sum rule.
The $\alpha_s$ corrections are given by diagrams of Fig.\ \ref{fig:gr4} and 
contain one gluon exchange. We calculate the expressions corresponding to
 all six diagrams (Fig.\ \ref{fig:gr4})  at zero momenta
 transfer. 
Subsequently, we perform the substitution (\ref{h9}) and retain only the 
terms
containing the structure (\ref{h14}). The contribution of each diagram has 
the same structure:
\begin{equation}
-\phi _i\frac{g^2}{(4\pi)^4}\log[\frac{p^2}{4\pi \mu ^2}]
\frac{q^{\lambda}A_{\lambda}}
{q^2}\epsilon_{\mu\nu\alpha\beta}q^{\alpha}p^{\beta} 
\label{h16}
\end{equation}
\begin{eqnarray}
\alpha_s=\frac{g^2}{(4\pi)} \nonumber
\end{eqnarray}

The obtained numerical factors $\phi_i$'s are listed below:
\begin{eqnarray}
\phi _a+\phi _b=-i\frac{16}{3} ;\ \ \ \ 
\phi _c=i\frac{16}{9};\nonumber \\  \phi _d=i\frac{4}{9};\ \ \ \ 
\phi _e+\phi _f=i\frac{56}{9}; \nonumber
\end{eqnarray}
The appearance of the $\log$ term, of course, does not surprise. 
What is less obvious is a complete cancellation
of all infrared divergencies that  occur in  the mass-less limit
we are working in. This remarkable fact signifies  the existence of 
OPE and the fact that long and short distances are  separated indeed.
Taking the sum, we obtain that  the  contribution of the unit operator to 
OPE is given by: 
\begin{equation}
-i\frac{28}{9(4\pi)^3}\alpha _s(p^2)\log[\frac{p^2}{4\pi \mu ^2}]
\frac{q^{\lambda}A_{\lambda}}
{q^2}\epsilon_{\mu\nu\alpha\beta}q^{\alpha}p^{\beta} 
\label{h17}
\end{equation} 
Here the running coupling constant $\alpha _s(p^2)$ is given by:
\begin{equation}
\alpha _s(p^2)=\frac{4\pi}{9\log[\frac{p^2}{\Lambda ^2}]}
\label{h18}
\end{equation}
$\Lambda$ denotes the cutoff of QCD and is taken to be 0.2 Gev.
\vskip10pt

Next, let us calculate the power corrections to the sum rule. We take
into account $\langle :\bar{\psi}\psi :\rangle $ and 
$\langle :\bar{\psi}\psi :\rangle _A$ condensates only.
The contributions due to $\langle :GG:\rangle $, $\langle :GG:\rangle _A$, 
$\langle :\bar{\psi}G\psi :\rangle $,
$\langle :\bar{\psi}G\psi :\rangle _A$ are negligible and the operators 
with $d\ge6$ are
omitted as well. We work in the leading order in $\alpha_s$.

The power corrections are given by the  diagrams of the
Fig.\ \ref{fig:gr5}, Fig.\ \ref{fig:gr6}.

Odd number of $\gamma$ matrices entering into the expression for the 
diagram Fig.\ \ref{fig:gr5} forces it to be zero.

The expression corresponding to the diagram Fig.\ \ref{fig:gr6} looks like:

\begin{eqnarray}
\langle  :\psi ^b(0)\bar{\psi}^a(x) 
:\rangle _A\gamma^{\mu}& &S_0(x)\gamma^{\nu}
\delta^{ab} \nonumber \\
-S_0(x)& &\gamma^{\mu}\langle :{\psi}^a(x)\bar\psi ^b(0) 
:\rangle _A\gamma^{\nu}\delta^{ab}
\nonumber
\end{eqnarray}

Here $S_0$ denote the standard quark propagator in $x$ representation. 
We use  (\ref{h8}) for the  condensates.
Performing the substitution (\ref{h9}) and retaining only terms containing
the structure  (\ref{h14}) we obtain the condensate contribution to  OPE.
It has the form:

\begin{equation}
i[\frac{f_\pi^2}{3p^2}-\frac{5}{27}\frac{m_1^2f_\pi^2}{p^4}]
\frac{q^{\lambda}A_{\lambda}}
{q^2}\epsilon_{\mu\nu\alpha\beta}q^{\alpha}p^{\beta} 
\label{h19}
\end{equation}

Now in order  to obtain the desired expression for
 $g_{\omega\rho\pi}$ it is a 
good time  to consider the sum rule. On the left hand side we have the 
expression (\ref{h13}), while  (\ref{h17}) + (\ref{h19}) stands on the 
right hand side.

The obtained sum rule reads:
\begin{eqnarray}
g_{\omega\rho\pi} \frac{m_{\rho}^2 m_{\omega}^2}{\sqrt 2g_\rho
g_\omega} \frac{f_\pi}{(p^2-m_{\rho}^2)^2} +...& \nonumber \\
&\nonumber \\
=
\frac{5}{27}\frac{m_1^2f_\pi^2}
{p^4}-\frac{f_\pi^2}{3p^2}+\frac{28}{9(4\pi)^3}&\alpha _s(p^2)\log[p^2/
4\pi \mu ^2]
\label{h20}
\end{eqnarray}

In order to suppress the contributions of higher resonances it is essential
to perform the Borel transformation \cite{SVZ79}. We introduce the continuum
 threshold $S_0$ in a standard way and transfer the continuum
contribution into the theoretical part of the sum rule. The integration
in the dispersion representation  of the unit operator is cut to the
integral from 0 to $S_0$. (Other diagrams  have no imaginary part).

Finally, we take into account the anomalous dimensions of the operators
\cite{BI82}
and obtain the following sum rule:
\begin{eqnarray}
 [g_{\omega\rho\pi}+CM^2] \frac{f_\pi m_{\rho}^2 m_{\omega}^2}
{\sqrt 2g_\rho g_\omega}  e^{-\frac{m_{\rho}^2}{M^2}}
=\frac{5}{27}m_1^2f_\pi^2 L^\gamma + \frac{f_\pi^2}{3}M^2 \nonumber \\ 
\nonumber \\ +
\frac{28}{9(4\pi)^3}   \alpha _s(M^2)M^4(1-
e^{-S_0/M^2})  
\label{h22} 
\end{eqnarray}
The piece proportional to $C$ takes into account the non-diagonal transitions
given by Fig.\ \ref{fig:gr2}. 
Here 

\centerline{$L=\frac{\log [M^2/\Lambda ^2]}{\log [\mu ^2/\Lambda^2]}; \ \ \ 
\gamma=-4/9;
\ \ \ \ \mu =0.5$ Gev.}  

For $m_{\rho}^2$ and the form-factors the following values are taken:

\centerline{
\ \ \ $m_{\rho}^2=0.59$ Gev${}^2$;\ \ \ $\frac{4\pi}{g_\rho}=0.41;\ \ \ 
\frac{4\pi}{g_\omega}=0.046$.}

We analyse the obtained sum rule.
We find the region of stability, where the contribution of the continuum is 
less than $30\%$. In order to have a right hierarchy of the power corrections
 we look
for a region, where the first term of the right-hand side 
 of (\ref{h22}) is also less than $30\%$.

In order to extract  $g_{\omega\rho\pi}$ from the obtained sum rule, we
transform  the equation  (\ref{h22}) and plot the 
function  $R(M^2,S_0)=g_{\omega\rho\pi}+CM^2$ for various values of $S_0$.
The desired  value for $g_{\omega\rho\pi}$ is given by the point where the 
straight lines 
asymptotically cross the $y$-axis (Fig.\ \ref{fig:gr10}).
Investigating the graph Fig.\ \ref{fig:gr10} we see that the region of 
stability begins
at $M^2 \simeq 1.2$ Gev$^2$. The determined value for $g_{\omega\rho\pi}$:

\centerline{$g_{\omega\rho\pi}\sim 16 $ Gev $^{-1}$.} 

The inaccuracy of the obtained value is about $10\%$.
The result does not change significantly with the variation
of $S_0$ that implies an insensitivity to the continuum threshold. Note also
that the $\alpha_s$ corrections
seem to be unimportant and the main contribution to the sum rule comes from
 the condensate terms. 

\section{conclusions}
\label{sec:con}
In the present work we investigated the constant of the
$ \omega\rightarrow\rho\pi$ transition by means of the QCD sum rules 
in external
field. As was expected, at small momentum transfer the method of external 
field works perfectly and provides us with a reliable answer. No instability
of the sum rules has occurred. The stability region and the optimal value
for the parameter of continuum  almost coincide with the values, which were
used to determine the mass of $\rho$ meson \cite{SVZ79}.

The obtained value for $g_{\omega\rho\pi}$ is about $16$ Gev $^{-1}$ with the 
relative error of about $10\%$. Our result  is in a good agreement with 
the experimental data, confirms the calculations of 
\cite{EK83,RRY83,NP84,MS88} and disproves the results of Ref. \cite{Kh85}.

\acknowledgements
I am deeply grateful to my supervisor Prof. Boris Blok  for the suggestion 
of the problem and the permanent guidence  during  the  work.   


\begin{figure}
\caption{The $\omega\rightarrow\rho$ transition in the presence of  external
field.}
\label{fig:gr1}
\end{figure}

\begin{figure}
\caption{The parasitic contributions to the 
$\omega\rightarrow\rho$ transition.}
\label{fig:gr2}
\end{figure}

\begin{figure}
\caption{Contribution of the unit operator to the polarisation operator, of
zero order in $\alpha_s$.}
\label{fig:gr3}
\end{figure}

\begin{figure}
\caption{Contribution of the unit operator to  the polarisation operator,
of the first order in $\alpha_s$.}
\label{fig:gr4}
\end{figure}

\begin{figure}
\caption{$\langle :\bar{\psi}\psi :\rangle $ contribution to the polarisation
operator.}
\label{fig:gr5}
\end{figure}

\begin{figure}
\caption{$\langle  :\bar{\psi}\psi :\rangle _A$ contribution to the 
polarisation operator.}
\label{fig:gr6}
\end{figure}

\begin{figure}
\caption{The value for $g_{\omega\rho\pi}$ from the sum rule. The cluster of
the six curves corresponds to the values of $S_0$ from 0 to 2.5. 
The outstanding curve corresponds to $S_0\rightarrow\infty$. }
\label{fig:gr10}
\end{figure}

\end{document}